\documentclass[aps,prb,preprint,groupedaddress,showpacs,showkeys]{revtex4-1}
\usepackage{graphicx}
\newcommand{\eqlabel}[1]{\label{#1}} 
\newcommand{\chem}[1]{$\mathrm{#1}$ } 

\newcommand{\NiMnO}{\chem{Ni_{0.5}Mn_{1.5}O_4}}
\newcommand{\LiNiMnO}{\chem{LiNi_{0.5}Mn_{1.5}O_4}}
\newcommand{\LixNiMnO}{\chem{Li_{\mathit{x}}Ni_{0.5}Mn_{1.5}O_{4}}}

\newcommand{\LixX}{\chem{Li_{\mathit{x}}\mathit{X}}}
\newcommand{\LihNiMnO}{\chem{Li_{0.5}Ni_{0.5}Mn_{1.5}O_{4}}}

\newcommand{\bl}{\bullet}
\newtoks\tmpId

\begin{document}

\title{Thermodynamic Analysis Using First-Principles Calculations
  of Phases and Structures
  of $\mathbf{Li{_x}Ni_{0.5}Mn_{1.5}O_{4} (0 \le x \le 1)}$
}

\author{Ippei~Kishida}
\email[]{kishida@imat.eng.osaka-cu.ac.jp}
\affiliation{
  Department of Mechanical Engineering,
  Faculty of Engineering,
  Osaka City University,
  3-3-138 Sugimoto, Sumiyoshi-ku, Osaka 558-8585, Japan
}
\author{Kengo~Orita}
\affiliation{
  Department of Mechanical Engineering,
  Faculty of Engineering,
  Osaka City University,
  3-3-138 Sugimoto, Sumiyoshi-ku, Osaka 558-8585, Japan
}
\author{Atsutomo~Nakamura}
\affiliation{
  Department of Mechanical Engineering,
  Faculty of Engineering,
  Osaka City University,
  3-3-138 Sugimoto, Sumiyoshi-ku, Osaka 558-8585, Japan
}
\author{Yoshiyuki~Yokogawa}
\affiliation{
  Department of Mechanical Engineering,
  Faculty of Engineering,
  Osaka City University,
  3-3-138 Sugimoto, Sumiyoshi-ku, Osaka 558-8585, Japan
}

\date{\today}

\begin{abstract}
\LiNiMnO, which has a spinel framework structure,
is a promising candidate for the cathode material of next-generation lithium-ion batteries
with high energy density.
We investigate the structural transition in \LixNiMnO ($0 \le x \le 1$)
through first-principles calculations using the projector augmented wave
method with the generalized gradient approximation.
We calculate all the unique Li-site occupation configurations
in a unit cell 
to obtain the
total energies and the most stable structures
for various compositions.
Thermodynamic analysis shows that
\LihNiMnO with $x = 0.5$ is the only stable phase
for $0 < x < 1$.
The decomposition energy is lower than 0.1 eV
for $0 < x < 0.5$,
but is distinctly higher for $0.5 < x < 1$. The decomposition energy
reaches 0.39 eV at $x=0.75$.
The ratios of the structures at room temperature are calculated
from Boltzmann factors by using the energy differences between structures.
The crystal structure of the unit cell changes gradually
from $x=0$ to $0.5$,
but changes markedly from $x=0.5$ to $1$.
This first-principles study provides
a general evaluation of
the variation in the crystal structure with the composition of the bulk material,
which affects the cyclability of the electrode.
\end{abstract}


\maketitle

\section{Introduction}
the high energy density of lithium-ion batteries 
makes them particularly suitable energy storage devices for the many new portable electronic devices that are coming onto the market.
\chem{LiCoO_2} is widely used as a cathode material for lithium-ion batteries,
because it has a voltage of up to 4~V versus metallic lithium
\cite{Mizushima1980}.
Voltage is a fundamental property of batteries;
higher voltages are closely related to 
larger electrical energy capacities
and faster electronic performance in processing units.
Recently,
several transition-metal-doped spinel cathode materials with voltages of 5~V, such as \chem{LiM_{\mathit{x}}Mn_{2-\mathit{x}}O_{4}} (M = Ni, Cr, Co, Fe, Cu),
have been reported \cite{Koksbang1996, Xu2010, Churikov2011, Cho2011}.
\LiNiMnO is a particularly promising material
because of its high voltage, electrochemical stability,
and lack of trace metal impurities
\cite{Zhong1997, Fang2006}.
Furthermore, materials such as \chem{ Li[MnNiCo]O_{2} } \cite{Wu2011, Cho2011},
\chem{ Li[MnNiCr]O_{2} } \cite{Karan2009},
and \chem{ Li[MnNiFeCo]O_{2} } \cite{Son2007} are based on \LiNiMnO.
The physical and chemical properties
of \LiNiMnO have been experimentally determined by X-ray diffraction, scanning electron microscopy,
infrared-Raman spectroscopy, electron diffraction,
and cyclic voltammetry 
\cite{Ariyoshi2004, Mat2006, Xia2007}.
Although the crystallographic and electrochemical properties of \LiNiMnO have been well reported,
the microscopic changes in the crystal structure and electrochemical properties
during the lithiation and delithiation of an \LixNiMnO crystal
are not yet fully understood.
Charging and discharging are primary functions of secondary batteries;
therefore, the change in the composition and structure of electrodes is unavoidable, and
the changes influence cyclability.
Predicting these structural changes will help to accelerate the development of electrode materials.

Theoretical and microstructural analysis can provide a valuable alternative to experimental studies for understanding the properties of materials.
First-principles calculation based on
density functional theory (DFT) is among the most powerful methods
for determining the fundamental characteristics of electrodes.
The theoretical band structures and crystal structures of
various Li-transition metal oxides,
including \chem{LiMnO_2} \cite{Aydinol1997b, Koyama2002} and
\chem{LiMn_2O_4} \cite{Koyama2003}, have been thoroughly investigated.
Several studies using theoretical calculations and basic thermodynamic models
have shown that the
average electrochemical potential of \LiNiMnO is derived from the
difference in total energy between the fully lithiated and delithiated materials
\cite{Aydinol1997a, Aydinol1997b, Yi2008}.
However, a method has not yet been established for evaluating the changes in crystal structure
and electrochemical potential
with varying Li composition.
To analyze the properties of crystals at intermediate compositions quantitatively,
we investigated the sublattice structures of Li 
and used DFT to evaluate their thermodynamic stability
at various compositions between \NiMnO and \LiNiMnO.
We propose a method for predicting 
the average electrochemical potential
as well as the change in electrochemical potential as the composition
of the bulk material is varied.
This information is useful
for designing new electrode materials,
because the electrochemical potential and cyclability
of an electrode material is strongly affected by
the change in the atomic structure
during lithiation and delithiation.

\section{Computational Procedure}
\subsection{Crystal Structure}
The starting structure for our study was a unit cell of \LiNiMnO crystal (Fig. \ref{fig20121219a}).
The cell belongs to the \chem{P4_{3}32} space group
and consists of 8 formula units containing 56 atoms in total:
\chem{Li_8Ni_4Mn_{12}O_{32}}
\cite{Ariyoshi2004}.
The O atoms are located at the 8(c) and 24(e) sites,
and form a distorted face-centered cubic (fcc)
structure in the cell.
Cations occupy the ordered polyhedral sites formed by the O sublattice;
the Li, Ni, and Mn atoms are present in the octahedral 8(c) sites,
4(b) sites, and 12(d) sites, respectively.
The Li sublattice forms a slightly distorted diamond structure.
The eight Li sites in the unit cell
were labeled as shown in Table \ref{table20120316b}.
The initial cells for geometry optimization at various compositions
were made by removing $n$ ($n = 0, 1, 2, \cdots, 8$) Li atoms
from the \chem{Li_{8}Ni_{4}Mn_{12}O_{32}} cell,
resulting in a \chem{Li_{8-\mathit{n}}Ni_4Mn_{12}O_{32}} cell.
The Li sites are symmetrically equivalent in the space group.
Therefore, for $n = 0, 1, 7$, and $8$, the unit cell crystal structure is the same.
For $2 \le n \le 6$,
the configurations of the occupied and unoccupied Li sites were considered.
We calculated all the unique configurations in the symmetry operations
of its space group (Table \ref{table20120720a}).
The geometry of each cell was optimized 
to obtain the most stable lattice volume, shape, and atomic coordinates;
the total energies of the cells were computed at the same time.
We examined differences in energy of the various structures.
The change in Gibbs energy is expressed by
  $\Delta G = \Delta E + P \Delta V - T \Delta S.$
However,
$\Delta E$ was of the order of 0.04-4 (eV/cell),
$P \Delta V$ was of the order of $10^{-5}$ (eV/cell),
and $T \Delta S$ was of the order of the thermal energy in this study
\cite{Aydinol1997, Aydinol1997b, Courtney1998}.
Because the $P \Delta V$ and $T \Delta S$ terms have little effect on $\Delta G$,
the assumption
that $\Delta G$ can be approximated by only the change in the internal energy,
$\Delta E$,
is valid.

\begin{figure}
  \tmpId={fig20121219a}
  \begin{center}
  \includegraphics[width=1.0\linewidth]{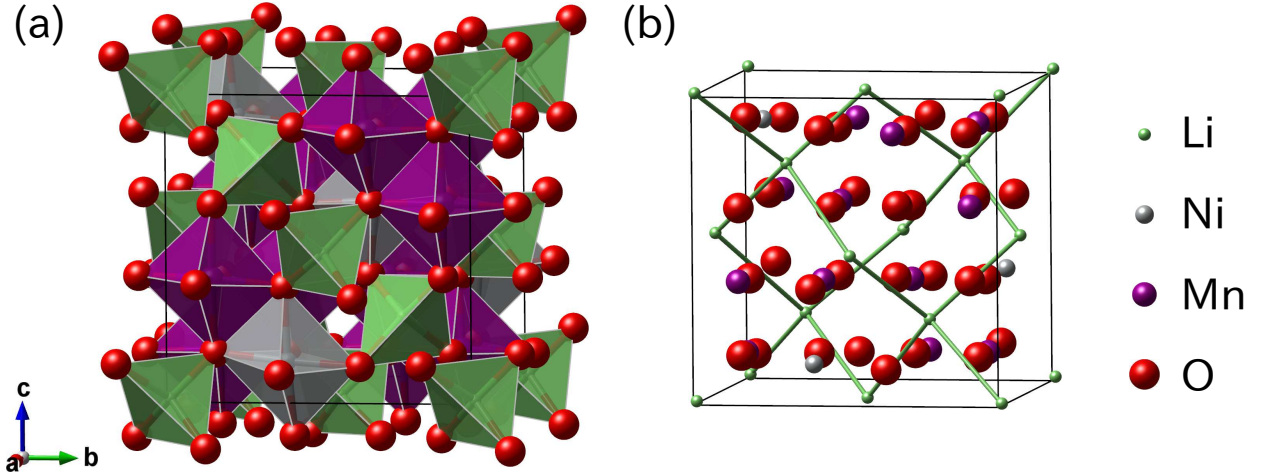}
  \end{center}
  \caption{ 
    \label{\the\tmpId}
    (Color online) 
    Crystal structure of \LiNiMnO.
    The Li, Ni, Mn, and O ions are shown in green, gray, purple, and red, respectively.
    The radii are different to allow them to be clearly distinguished:
    $r_{\mathrm{Li}} < r_{\mathrm{Ni}} < r_{\mathrm{Mn}} < r_{\mathrm{O}}$.
    (a)  O polyhedra including cations. The tetrahedra (green),
    light gray octahedra, and dark gray octahedra (purple) indicate the 
    \chem{LiO_4}, \chem{NiO_6}, and \chem{MnO_6} polyhedra, respectively.
    (b) Li-Li bonding in the Li sublattice.
  }
\end{figure}

\begin{table}
  \tmpId={table20120316b}
  \caption{ 
    \label{\the\tmpId}
    Labels and internal coordinates of Li sites in the unit cell.
    These coordinates are adjusted to those of the ideal diamond structure to allow for easy        
    recognition.
  }
  \begin{center}
  \begin{tabular}{cllllllll}
  \hline 
  \hline 
  axis & L000& L220& L111& L331& L202& L022& L313& L133 \\
  \hline 
  $a$  &0   & 0.5 & 0.25& 0.75& 0.5 & 0   & 0.75& 0.25\\
  $b$  &0   & 0.5 & 0.25& 0.75& 0   & 0.5 & 0.25& 0.75\\
  $c$  &0   & 0   & 0.25& 0.25& 0.5 & 0.5 & 0.75& 0.75\\
  \hline 
  \hline 
  \end{tabular}
  \end{center}
\end{table}

\begin{table}
  \tmpId={table20120720a}
  \caption{ 
    \label{\the\tmpId}
    Unique configurations of
    occupied and unoccupied Li sites
    in a unit cell of \chem{Li_{\mathit{n}}Ni_{4}Mn_{12}O_{32}}.
    All configurations are assigned an uppercase character from A to W.
    The filled circles and dashes indicate occupied and unoccupied sites,
    respectively.
    The labels at the top, for example L000, are site names,
    which are the same as those in Table~\ref{table20120316b}.
  }
  \begin{center}
  \begin{tabular}{ccllllllll}
  \hline 
  \hline 
  $n$& structure& L000& L220& L111& L331& L202& L022& L313& L133 \\
  \hline 
  0&   A&   --&   --&   --&   --&   --&   --&   --&   -- \\
  \hline 
  1&   B&$\bl$&   --&   --&   --&   --&   --&   --&   -- \\
  \hline 
  2&   C&$\bl$&$\bl$&   --&   --&   --&   --&   --&   -- \\
  2&   D&$\bl$&   --&$\bl$&   --&   --&   --&   --&   -- \\
  2&   E&$\bl$&   --&   --&$\bl$&   --&   --&   --&   -- \\
  \hline 
  3&   F&$\bl$&$\bl$&$\bl$&   --&   --&   --&   --&   -- \\
  3&   G&$\bl$&$\bl$&   --&$\bl$&   --&   --&   --&   -- \\
  3&   H&$\bl$&$\bl$&   --&   --&$\bl$&   --&   --&   -- \\
  \hline 
  4&   I&$\bl$&$\bl$&$\bl$&$\bl$&   --&   --&   --&   -- \\
  4&   J&$\bl$&$\bl$&$\bl$&   --&$\bl$&   --&   --&   -- \\
  4&   K&$\bl$&$\bl$&$\bl$&   --&   --&   --&$\bl$&   -- \\
  4&   L&$\bl$&$\bl$&$\bl$&   --&   --&   --&   --&$\bl$ \\
  4&   M&$\bl$&$\bl$&   --&$\bl$&$\bl$&   --&   --&   -- \\
  4&   N&$\bl$&$\bl$&   --&$\bl$&   --&   --&   --&$\bl$ \\
  4&   O&$\bl$&$\bl$&   --&   --&$\bl$&$\bl$&   --&   -- \\
  \hline 
  5&   P&$\bl$&$\bl$&$\bl$&$\bl$&$\bl$&   --&   --&   -- \\
  5&   Q&$\bl$&$\bl$&$\bl$&$\bl$&   --&$\bl$&   --&   -- \\
  5&   R&$\bl$&$\bl$&$\bl$&   --&$\bl$&$\bl$&   --&   -- \\
  \hline 
  6&   S&$\bl$&$\bl$&$\bl$&$\bl$&$\bl$&$\bl$&   --&   -- \\
  6&   T&$\bl$&$\bl$&$\bl$&$\bl$&$\bl$&   --&$\bl$&   -- \\
  6&   U&$\bl$&$\bl$&$\bl$&$\bl$&   --&$\bl$&$\bl$&   -- \\
  \hline 
  7&   V&$\bl$&$\bl$&$\bl$&$\bl$&$\bl$&$\bl$&$\bl$&   -- \\
  \hline 
  8&   W&$\bl$&$\bl$&$\bl$&$\bl$&$\bl$&$\bl$&$\bl$&$\bl$ \\
  \hline 
  \hline 
  \end{tabular}
  \end{center}
\end{table}

\subsection{Thermodynamic analysis of phases}
The ratio of structures at equilibrium was determined for
a total Li content in the bulk of $x=b$.
The lithiated and delithiated material was assumed to be \LixX,
where $X$ was \NiMnO.
Because this is a pseudo-unitary system of Li,
we examine Li content only.
For compositions of $x = a, b$, and $c~(a < b < c)$ in
\LixX,
the decomposition reaction is
\begin{eqnarray}
  \mathrm{Li_{\mathit{b}}}X
  \rightarrow
  \frac{c-b}{c-a} \mathrm{Li_{\mathit{a}}}X
  +
  \frac{b-a}{c-a} \mathrm{Li_{\mathit{c}}}X
  . \eqlabel{eq20120315c}
\end{eqnarray}
The decomposition energy of this reaction,
$\Delta G_b(a, c)$,
is
\begin{eqnarray*}
  \Delta G_b(a, c)
  &=& G(b) - \left( \frac{c-b}{c-a} G(a) + \frac{b-a}{c-a} G(c) \right)
  ,  \eqlabel{eq20120315b}
\end{eqnarray*}
where $G(x)$ is the Gibbs energy of the most stable structure at composition $x$.
Positive values of $\Delta G_b(a, c)$ indicate
that the left side of Eq. (\ref{eq20120315c})
is thermodynamically unstable, whereas negative values indicate that it is stable.
$\Delta G_b(a, c)$ gives
the ratio of the equilibrium amounts of species on each side of Eq. (\ref{eq20120315c})
by using the Boltzmann factor, 
\begin{eqnarray}
  \frac{ P_b(a, c) }{ P_{b}(b) }
  &=&
  \exp \left(
    \frac{\Delta G_b(a, c)}{kT}
  \right),
  \eqlabel{eq20120930a}
\end{eqnarray}
where
$P_{b}(b)$ and $P_{b}(a, c)$
are the ratios of both edges in Eq. (\ref{eq20120315c}) in the bulk,
$k$ is the Boltzmann constant, and $T$ is the temperature.
$T$ was assumed to be room temperature, 298~K.
The sum of the ratios of all states
 at a given a given value of $b$
must be 1:
\begin{eqnarray}
  P_b(b) + \sum_{0 \le a < b} ~ \sum_{ b < c \le 1} \Big( P_{b}(a, c) \Big) &=& 1.
  \eqlabel{eq20121001c}
\end{eqnarray}
The composition was treated as a discrete value.
Each ratio of the decomposed state in the bulk,
$P_b(a,c)$ and $P_b(b)$, was calculated by solving simultaneous equations
\ref{eq20120930a} and \ref{eq20121001c}.
Then, each ratio, $P_{b}(x)$, was derived from
\begin{eqnarray}
  P_{b}(x) =
  \left\{
    \begin{array}{ll}
      \sum_{b < c \le 1} \frac{c-b}{c-x} P_{b}(x, c) &(x < b)
      \\
      P_{b}(b) &(x = b)
      \\
      \sum_{ 0 \le a < b} \frac{b-a}{x-a} P_{b}(a, x) &(x > b) .
    \end{array}
  \right.
  \eqlabel{eq20121001d}
\end{eqnarray}

Assuming that ideal monovalent \chem{Li^{+}} cations are the only charge carriers,
the average electrochemical potential versus \chem{Li/Li^+}
between $x_1 < x <x_2$ of \LixX, $V(x)$,
is given by
\begin{eqnarray}
  V (x) &=&
  - \frac{1}{F}
  \left(
    \frac{G(x_2) - G(x_1)}{x_2 - x_1}
  - G_{\mathrm{Li}}
  \right),
  \eqlabel{eq20120919a}
\end{eqnarray}
where $F$ is the Faraday constant,
and $G_{\mathrm{Li}}$ is the chemical potential of Li in a metallic Li anode
\cite{Aydinol1997, Aydinol1997b, Chevrier2010}.

\subsection{Computational conditions}
The spin-polarized calculations were performed by DFT
\cite{Hohenberg1964,Kohn1965} 
using the plane-wave projector augmented-wave method
\cite{Blochl1994} 
as implemented in the VASP code.
\cite{Kresse1993, Kresse1996, Kresse1999} 
The exchange-correlation term was treated
with the Perdew-Burke-Ernzerhof functional \cite{Perdew1996}. 
For the unit cell of \LixNiMnO, a cutoff energy of 500~eV and
a $k$-mesh of $4 \times 4 \times 4$
were determined through preliminary test calculations.
A cutoff energy of 500~eV and a $k$-mesh of $64 \times 64 \times 64$ were calculated for Li metal in a primitive cell with body-centered cubic structure.
Geometry optimization was truncated when the residual forces on the atoms became less than 0.02 eV/\AA.

\section{Results and discussion}
\subsection{\label{sec20121203a} Thermodynamic stability and phase transition}
Fig. \ref{fig20120316a} shows
the calculated energies of all the structures after geometry optimization.
For clarity, the values of energies shown on the vertical axis, $G'(x)$,
were corrected with
linearly approximated chemical potential of Li,
$\mu_{\mathrm{Li}}'$:
\begin{eqnarray}
  \mu_{\mathrm{Li}}' &=& G(1) - G(0)
  \eqlabel{eq20130116a}
  \\
  G'(x) &=& G(x) - (G(0) +  x \mu_{\mathrm{Li}}').
  \eqlabel{eq20121003b}
\end{eqnarray}
The line connecting the energies of the most stable structures for neighboring
compositions
was concave except for the area around $x = 0.5$ (Fig. \ref{fig20120316a}).
This indicates that only one composition, $x = 0.5$,
was thermodynamically stable, except for the end member
compositions, $x=0$ and $1$.
The Li sublattice in the cell labeled O
forms an fcc-like structure.
The fcc structure has high symmetry and uniform distance
between atoms.
The electrostatic repulsion between \chem{Li^{+}} ions makes it preferable for them to be positioned as far apart as possible. Thus, the high stability of the cell is reasonable.
In the region $0 < x < 0.5$,
the energy profile was concave, suggesting that \LixNiMnO should decompose to
\NiMnO and \LihNiMnO at 0 K.
However, the decomposition energy, $\Delta E_x(0, 0.5)$,
was less than 0.1~eV.
This slight instability would produce the structures
labeled B, C, and H at intermediate compositions 
through the thermal effect, even at room temperature.
The most stable structure changes were
A, B, C, H, and O for compositions varying from $x=0$ to $x= 0.5$.
Table \ref{table20120720a} shows that the change in the most stable structure between neighboring compositions can be represented by replacing
occupied Li sites with unoccupied sites,
or unoccupied sites with occupied sites. %
For example,
the only difference between structures H and O is at the L022 site.
This simple mechanism would result in smooth changes
in the atomic structure as the composition changes
and may contribute to good cyclability.
In contrast,
the decomposition energy, $\Delta E_x(0.5, 1)$, was 
high in the region $0.5 < x \le 1$.
In particular, $\Delta E_{0.75}(0.5, 1)$ was 0.39~eV for structure S.
The bulk material should be a mixture of two separate phases, namely,
structures O and W.
Therefore, the crystal structure of the unit cell changed considerably in this region.
The cells that transformed from \LihNiMnO to \LiNiMnO were stressed by
the volume change,
which was estimated to be about 4\%.

\begin{figure}
  \tmpId={fig20120316a}
  \begin{center}
  \includegraphics{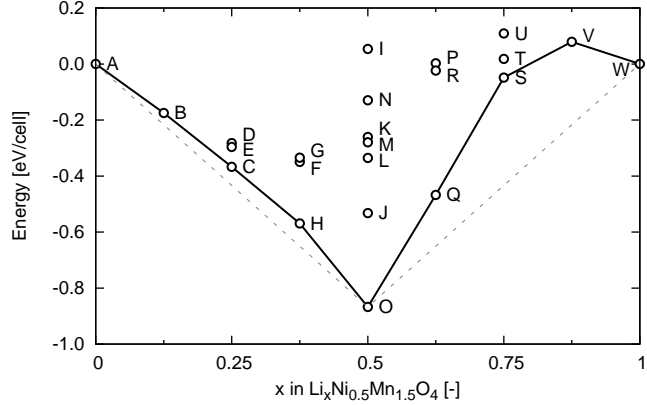} 
  \end{center}
  \caption{ 
    \label{\the\tmpId}
    Energies of crystal structures with a composition of \LixNiMnO.
    Each energy on the vertical axis is corrected by
    linear approximation of the Li chemical potential 
(Eqs.
    \ref{eq20130116a} and
    \ref{eq20121003b}).
    Labels A to W correspond to those
    in Table \ref{table20120720a}.
    The solid line connects
    the most stable structures at the various compositions.
    The dotted line
    connects
    all the thermodynamically stable structures between \NiMnO and \LiNiMnO.
  }
\end{figure}

The most stable structures,
which are on the solid line in Fig. \ref{fig20120316a},
are dominant at certain concentrations
because of the Boltzmann factor.
Therefore,
we can assume that 
they are the only structures in the bulk material.
We used the energies in Fig. \ref{fig20120316a}
and Eq. \ref{eq20121001d} to calculate the
ratio of structures in bulk \LixNiMnO at 298~K (Fig. \ref{fig20120928a}).
This figure shows that
the unit cell structure changed gradually via intermediate concentrations
when the total Li content $x$ was varied from $0$ to  $0.5$,
whereas the structure changed markedly from $x= 0.5$ to $1$.
These results suggest that
it would be difficult to
detect a difference in the internal structure of \LixNiMnO ($0 \le x \le 0.5$)
by experimental methods.
The cyclability of an electrode is expected to depend on how smoothly its crystal structure changes.
Therefore, it is valuable
to describe the change in structure quantitatively as in Fig. \ref{fig20120928a}
in order to predict the durability
of new electrode materials.

\begin{figure}
  \tmpId={fig20120928a}
  \begin{center}
  \includegraphics{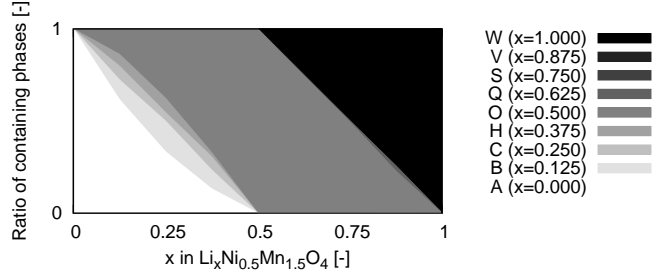} 
  \end{center}
  \caption{ 
    \label{\the\tmpId}
    Calculated ratio of structures in bulk \LixNiMnO at 298~K.
    The value of $x$ on the horizontal axis indicates the total Li content in the bulk.
    Darker areas indicate structures
    with higher Li concentration.
    The uppercase characters in the legend indicate structures
    corresponding to those in
    Table \ref{table20120720a} and Fig. \ref{fig20120316a}.
    The value of $x$ for each of the uppercase characters
    indicates the Li occupancy of the structures.
  }
\end{figure}

\subsection{Electrochemical potential}
There were three thermodynamically stable compositions, where $x = 0, 0.5$, and $1$. 
The electrochemical potential in the region $0 < x < 0.5$ and
$0.5 < x < 1$ was calculated by using Eq.~(\ref{eq20120919a}).
Fig. \ref{fig20120316c} shows the variation of the electrochemical potential
with the total Li content in a bulk electrode.
The calculated values of 3.99 and 3.56~V
were lower than the experimental values of 4.739 and 4.718~V
\cite{Ariyoshi2004}.
However, first-principles calculations
using the generalized gradient approximation  typically give underestimates
\cite{Aydinol1997b},
and the calculated and experimental results showed good agreement.
The total change in the theoretical electrochemical potentials
for compositions of $0 \le x \le 1$
was calculated as 0.43~V.
The changes in the electrochemical potential shown in Fig. \ref{fig20120316c}
correspond to the changes in the slope of the dotted line
in Fig. \ref{fig20120316a} because the electrochemical potential is reflected in $\mu_{\mathrm{Li}}$ of the material.
The small change in electrochemical potential resulted from
\LihNiMnO being only slightly stable, that is,
having slightly negative decomposition energy.
the small stability of \LihNiMnO.
If there were no stable phase in intermediate composition of \LixX,
the material would theoretically have constant electrochemical potential.
However, 
during lithiation and delithiation, an intermediate composition that is highly unstable results in
a sharp change in atomic structure, which is an obstacle to good cyclability.

\begin{figure}
  \tmpId={fig20120316c}
  \begin{center}
  \includegraphics{\the\tmpId.eps} %
  \end{center}
  \caption{ 
    \label{\the\tmpId}
    Calculated changes in the electrochemical potential of \LixNiMnO versus \chem{Li/Li^+}.
  }
\end{figure}

\section{Conclusion}
In summary,
first-principles calculations revealed the microscopic structure and stability
of \LiNiMnO,
which is among the most promising materials for lithium-ion battery cathodes.
We quantitatively evaluated the energies of all the configurations of Li-site occupation for intermediate compositions
of \LixNiMnO ($0 \le x \le 1$).
In this way, we assessed the thermodynamic stability of each structure and the change
in the electrochemical potential with varying total Li content in the bulk.
\LihNiMnO ($x = 0.5$) was the only stable phase for $0 < x < 1$.
The decomposition energy was lower than 0.1~eV for $0 < x < 0.5$,
but was high for $0.5 < x < 1$.
The thermodynamic analysis showed that the
change in structure with varying Li content was gradual for low Li content compositions,
and rapid for high Li content compositions.
The theoretical change in electrochemical potential at $x = 0.5$
was found to be 0.43~V, which reflects \LihNiMnO having lower stability than \NiMnO and \LiNiMnO.
We have demonstrated that theoretical calculations can 
reveal precise changes in structure, phase, and electrochemical potential
and can explain the cyclability of electrode materials.
In particular, visualizing the change in the crystal structure
as shown in Fig. \ref{fig20120928a} will be particularly valuable for discovering
durable electrode materials.

\begin{acknowledgments}
We thank Dr. K. Ariyoshi for fruitful discussion.
\end{acknowledgments}

%

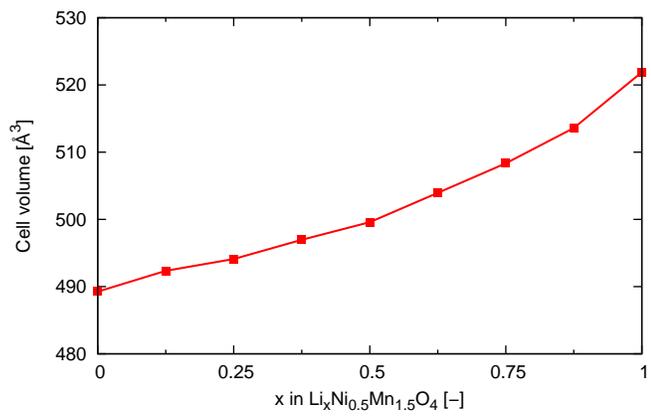
\end{document}